# ON NONLINEAR DYNAMICS OF A NON–IDEAL MAGNETIC SYSTEM WITH SHAPE MEMORY ALLOY TO ENERGY HARVESTING USING UNCERTAINTY EXPONENT AND ENTROPY OF BASIN OF ATTRACTIONS APPROACH,

M A RIBEIRO\*, J M BALTHAZAR\*\*  A M TUSSET \*  J J LIMA\*, \*\*\*\* , J L P FELIX\*\*\* ,V PICCIRILLO\*.

\*UTFPR-Federal University of Technology, Ponta Grossa, Paraná, Brazil

\*\* UNESP-State University of São Paulo, School of Engineering, São Paulo State University, Bauru, SP. Brazil

\*\*\*UFFS, Federal University of Fronteira Sul, Cerro Largo, Rio Grande do Sul, Brazil

\*\*\*\* University Visitor, University of Southampton, UK

**Abstract:**

In this paper, we dealt the dynamic behavior of a magnetic structure, considering a shape memory spring and a non-ideal motor as excitation. It has fractal behavior of tits basins of attraction, for a set of parameters for energy harvesting. We also considered a set of initial conditions and performed the nonlinear dynamics analysis of the dimensionless mathematical model and then established their chaotic and periodic regions. We also established the regions of maximum and minimum average power output, which was generated by the vibrations, caused by the considered non-ideal motor.

## 1- INTRODUCTION AND BRIEF COMMENTS ON THE STATE OF THE ART

In this paper, we examine a basic mathematical modeling assumptions for inertial, electric, and magnetic circuits, which are typical of mechatronic systems. We also summarize the dynamic principles and the interactions between the mechanical motion, circuit, and magnetic state variables. Usually, the governing equations of motion are in integral forms of the basic Partial Differential Equation (PDE) and may result in a coupled ordinary differential equation (ODEs). This methodology will be explored in this paper

[1]. By other hand, the study of systems which respond disproportionately (nonlinearly) to initial conditions is of great interest in Engineering and Science. Understanding of a nonlinear phenomenon in a variety of dynamical systems help us to deeply understand its various features and explore its applicability in real practical situations. The research development on nonlinear dynamics nowadays continuously reveals that nonlinear phenomena can bring many amazing and advantageous effects in very practical engineering problems such as vibration control, energy harvesting, structure health monitoring, micro/nano-electro-mechanical systems [1-10]

It is known that in recent years, the demand for energy consumption has been growing, so numerous devices have been designed to produce clean and sustainable energy. Examples of these devices are wind turbines that convert wind movement into electrical energy [2,3]. Another example of converting kinetic energy into electrical energy is ocean buoys; such devices convert the movement of ocean waves using portal frame structures with the coupling of piezo-ceramic materials [4-6], other ocean buoys use an electromagnetic piston and convert the oscillations of the waves into electrical energy.

Other devices consider a system based on the Duffing oscillator that considers a support with a beam crimped with a free end under the action of a magnetic potential formed by two magnetic poles. In this device, the beam has coupled piezo-ceramic materials that are coupled. When an external force is applied to the sinusoidal structure, energy is produced and collected by a circuit. In [7], the authors explore these devices with one and two degrees of freedom and investigate the non-linear dynamic behavior for energy production and the proposed parameters that can maximize energy production. Another factor explored by the authors is the diagnosis of chaos in structure and its effects on energy harvesting [9, 10].

Some devices that have gained some prominence for energy production are those that have SMA (Shape Memory Alloy) materials, and the outstanding feature of these materials is that after their deformation, they can return to their original shape when undergoing temperature variation. Works such as [8, 9] explore the non-linear dynamics of these structures by counting SMA for energy production using a Portal Frame type geometry considering the coupling of piezoceramic materials for energy collection. The authors propose not only a non-linear dynamic analysis of the energy harvesting system, but also a control project to suppress chaos for a set of parameters [11-28].

Therefore, analyzing the behavior of the non-linear dynamics of devices that produce energy for their collection is vital to improve their performance and thus propose parameter windows for periodic movements and possibly obtain more energetic orbits.

We remarked that, in the Design of structures, it is necessary to investigate the relevant dynamics to predict the structural response due to the excitations. In general, all the publications were based on the assumptions that the external excitations are produced by an ideal source of power with prescribed time history and prescribed magnitude, course, and frequency, or in random problem with prescribed characteristics. But the excitation sources are non-ideal, they have always limited power, limited inertia and their frequencies varies according to the instantaneous state of oscillating system [23]. In this point, it is important to clarify what rotary non-ideal systems means, in order, to avoid future confusions. Non-ideal systems (RNIS) have appeared in the literature with several meanings; as an example: some researchers use the concept of (RNIS) solutions for concentrated solutions, that is, the solutions can occur in two ways: when intermolecular forces between solute and solvent molecules are less strong than between molecules of similar (of the same type) molecules, and when intermolecular forces between dissimilar molecules are greater than those between similar molecules. Here, we deal with an energy transfer between the energy sources and the support structures and their possible control approaches, that is, we are interested to what happens to the DC motor input, output, as the response to the rotary system support structure changes [23-28]. Complete and comprehensive reviews of different theories on this subject can be found in [28-35] undeserved by others.

In this paper we, present some partial results at the EURODYN 2023 event as a short abstract and here we analyze the behavior of the nonlinear dynamics of an electromechanical system for energy harvesting. The device analyzed in this manuscript, is the one proposed before by [22], however, here, we consider an external force produced by a non-ideal motor.

Therefore, the aim of this paper is to analyze the nonlinear dynamic behavior of a device that converts the movement of a magnet of mass m, considering a spring with shape memory and a solenoid; in this structure we consider an external force exerted by a nonideal motor that will cause oscillations and as well as the movement of mass m producing an electric current. It is divided into the following sections: mathematical model, which describes the mathematical model and the scheme of the analyzed structure, the numerical results section that numerically explores the nonlinear dynamics with the basins of attraction analysis, maximum Lyapunov exponent, bifurcation diagram, phase maps and harvesting output potential.

## 2. OUTLINE OF THE MATHEMATICAL MODELING AND DISCUSSIONS

The energy harvesting device consists of a permanent magnet of mass *m*, supported by a spring of shape-memory material and *a* damping element *c*. The electrical circuit part is an RL element by introducing the principle of electromagnetic induction. The external force applied to the system is defined by a non-ideal motor that causes displacement *x,* the force is described by [10,29,35]:

$$F(t) = F_0 \cos[\Omega t + a_0 \sin(b_0 \Omega t)] \qquad (1)$$

where $F_0$ is the amplitude of the force acting on the system, $\Omega$ is a constant obtained from averaging of the angular frequency at resonance. The parameters $a_0$, $b_0$ and $c_0$ are defined by the active interaction between the oscillating system and the excitation source. In addition, they are considered control parameters. If $a_0 = 0$, then corresponding to harmonic excitation. Note that $a_0=0$ we have no action of the non-ideal motor, so:

$$cos[\Omega_0 t + a_0 \sin(b_0\Omega_0 t + c_0)] = \sum_{k=-\infty}^{\infty} J_k(a_0) \cos(\Omega_k t + kc_0),$$

$$\Omega_k = \Omega_0 + kb_0\Omega_0 \qquad (2)$$

In Fig. (1) (a) and (b) depicts the first four of these functions, which aid in understanding the effect of the nonideal excitation.

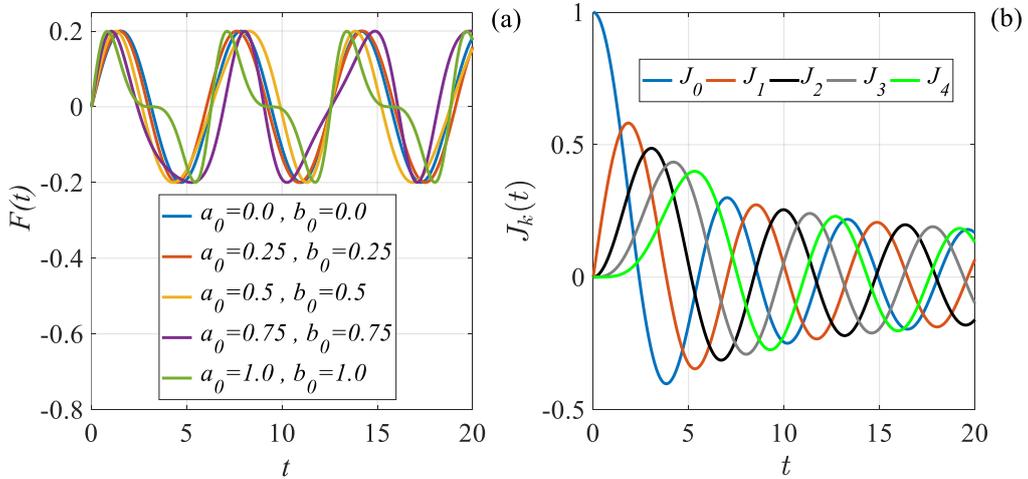

**Fig. (1):** (a) External force behavior for different values for *a₀* and *b₀* and (b) first four Bessel functions

The Shape Memory Alloy (SMA) action can be written as a polynomial describe in [22]:

$$F_s = \frac{aA}{L}(T - T_c)x - \frac{bA}{L^3}x^3 + \frac{eA}{L^5}x^5 \qquad (3)$$

where *a, b,* and *e* are the physical parameters of the materials, *T* is the temperature, $T_c$ is the critical deformation temperature, *A* denotes the area element of the spring. Then, the governing equations of motion of the device may be written as follows.

$$m\ddot{x} + c\dot{x} + F_s + BIL_c = F(t) \qquad (4)$$

$$L_i \dot{I} + RI - BL_c\dot{x} = 0$$

Where *m* is the mass of the permanent magnet, *x* is the response displacement, *c* is the damping coefficient, *L* the length of the spring, I the response current, $L_i$ the inductor, $L_c$ is the coil length, *F(t)* the applied external force. In Fig. 2(a) describes the mechanism and Fig. 2(b) the associated electrical circuit; the models were described by [16]:

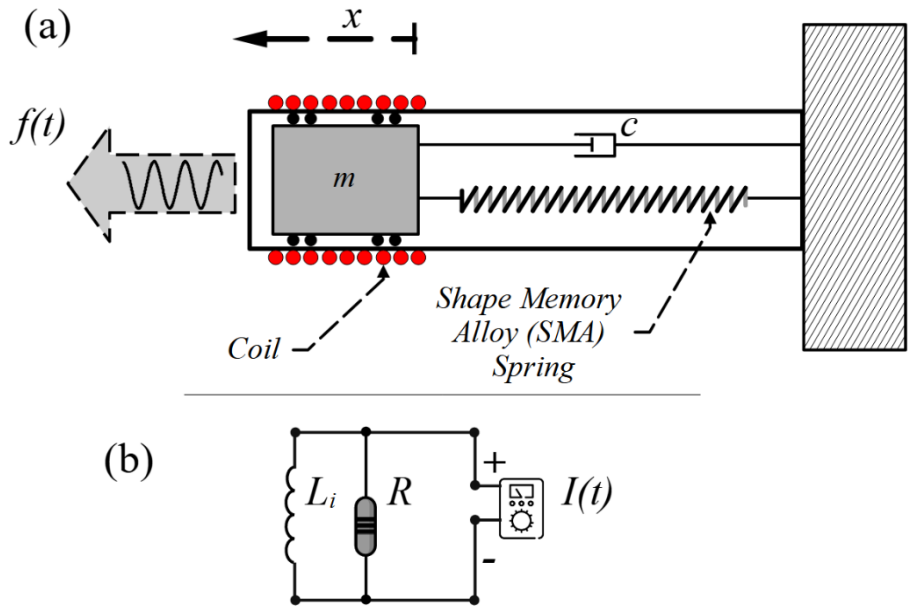

**Fig. (2): Simplify scheme of a shape memory vibratory energy harvester: (a) harvested device and (b) electric circuit [22].**

Therefore, we rewrite the governing equations of motion as follows:

$$m\ddot{x} + c\dot{x} + \frac{aA}{L}(T - T_c)x - \frac{bA}{L^3}x^3 + \frac{eA}{L^5}x^5 + BIL_c = F_0\cos[\omega t + a_0\sin(b_0\omega t + c_0)]$$

$$L_i\dot{I} + RI - BL_c\dot{x} = 0 \qquad (5)$$

Consider that $y = \frac{x}{L}$, $\tau = t\omega$, $\omega_0^2 = \frac{aAT_c}{mL}$, $\zeta = \frac{c}{m\omega_0}$, $\theta = \frac{T}{T_c}$, $\eta_1 = \frac{bA}{m\omega_0^2 L}$, $\eta_2 = \frac{eA}{m\omega_0^2 L}$, $y = \frac{x}{L}$, $f_0 = \frac{F_0}{m\omega_0^2 L}$, $v = \frac{L_i}{BL_c L}I$, $\beta = \frac{B^2 L_c^2}{m\omega_0^2 L_i}$ and $X = \frac{x}{\omega_0 L_i}$

Rewrite equations (5) in dimensionless equations, so:

$$\ddot{y} + \zeta\dot{y} + (\theta - 1)y - \eta_1 y^3 + \eta_2 y^5 + \beta v = F_0\cos[\omega t + a_0\sin(b_0\omega t + c_0)]$$

$$\dot{v} + Xv - \dot{y} = 0 \qquad (6)$$

Considering $x_1 = y$, $x_2 = \dot{y}$ e $x_3 = v$, so:

$$\dot{x}_1 = x_2$$

$$\dot{x}_2 = -\zeta x_2 - (\theta - 1)x_1 + \eta_1 x_1^3 - \eta_2 x_1^5 - \beta x_3 + f_0\cos[\omega t + a_0\sin(b_0\omega t + c_0)]$$

$$\dot{x}_3 = -Xx_3 + x_2 \qquad (7)$$

## 3. NUMERICAL RESULTS

The parameters used for the numerical analyzes are found in Ref. [22], however, we will list them in Table (1):

| Parameter | Value |
|---|---|
| $\zeta$ | 0.1 |
| $\eta_1$ | $1.3 \times 10^3$ |
| $\eta_2$ | $4.7 \times 10^5$ |
| $b_0$ | 1.0 |
| $\omega$ | 0.1492 |
| $\beta$ | 0.0113 |
| X | 0.0867 |
| $c_0$ | 0.0 |
| $\theta$ | 0.5 |

**Table (1): Parameter for numerical analysis**

Using the fourth order Runge-Kutta method and an integration step *h=0.01*, with a total integration time of $10^6$ [s] and a transient time of 40% of the total integration time. The first analysis performed was the behavior of the initial conditions of the system of Eqs. (5) [36,37,39].

### 3.1 BASINS OF ATTRACTION ANALYSIS

We observed the behavior of the basins of attraction of Eq. (7) for $x_1^0 \in [-0.08, 0.08] \times x_2^0 \in [-0.08, 0.08] \times x_3^0 = 0$ and the grid for plane $x_1 \times x_2$ contains $1200 \times 1200$ points. We consider $\theta = 0.5$, because according to [22] the potential has a double well behavior, which results in the bistability of the shape memory harvester. Therefore, for the analyzed range, four attractors were detected, for the parameters of Table (1).

We estimate the uncertainty exponent (α), that describes the sensitivity of the final state of the trajectories in the phase space. The algorithm used to determine the uncertainty exponent is the one proposed by [40], which randomly probes the basin of attraction with a ball of size ε. Thus, if there are at least two initial conditions that lead to different attractors, but belonging to the same ball of size ε, it will be marked as an uncertain ball. According to [41-42], the fraction of uncertain balls for the total number of trials in the basin of attraction, follow a power law ($f_\varepsilon \approx \varepsilon^\alpha$), the α that characterizes this scaling is called the uncertainty exponent. The exponent α close to 1 means that the basins of attraction have smooth contours, while a value close to 0 represents those basins of attraction are ridding.

The fig. (3) shows the uncertainty coefficient for the grid of initial conditions considered $x_1^0 \in [-0.085, 0.085] \times x_2^0 \in [-0.085, 0.085] \times x_3^0 = 0$, for Eqs. (7). The blue color shows values of α close to zero, which corresponds to the formation of sieve basins with the variation of the parameters *a₀* = [0, 0.5, 1.0] $f_0 \in [0.001, 0.04]$ which represent the parameters of the non-ideal engine considering Eqs. (7). However, for values of α close to 1 it shows the behavior of smoother basins of attraction and values close to 0 the basins of attraction have a sieve behavior. In Fig. (3) shows the behavior of the parameter α for Eqs. (7), in Fig. (3). a are the results obtained for *a₀* = 0 end *f₀* ∈

[0.001,0.04], Fig. (3). b $a_0 = 0.5$ end $f_0 \in$ [0.001,0.04] and Fig. (3) c $a_0 = 1.0$ and $f_0 \in$ [0.001,0.04]

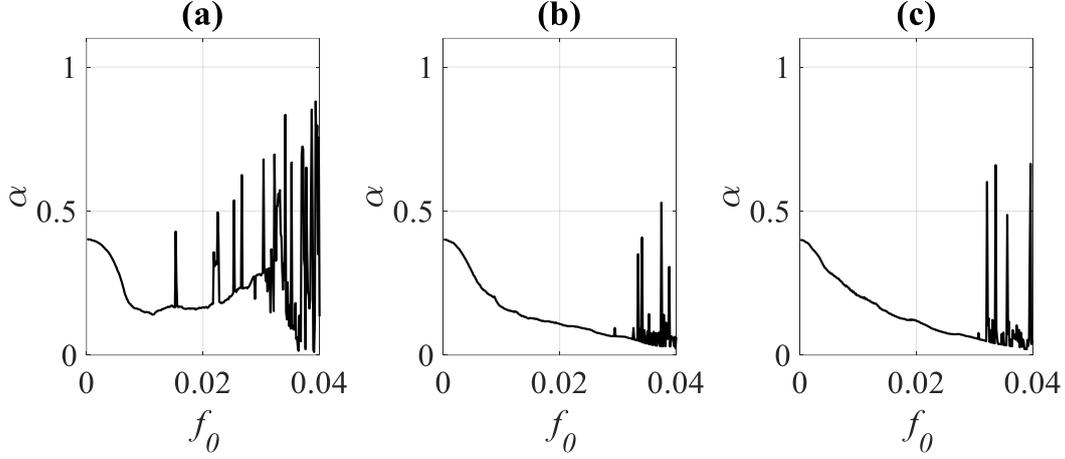

**Fig. (3):** Uncertainty exponent ($S_b$) for (a) $a_0 = 0.0$ and $f_0$ in [0.001, 0.04], (b) $a_0 = 0.5$ and $f_0$ in [0.001, 0.04] and (c) $a_0 = 1.0$ and $f_0$ in [0.001, 0.04].

To observe the sensitivity of the basins of attraction and to make a comparison with the exponent of uncertainty, we calculate the entropy of the basin of attraction ($S_b$). The entropy of the basins of attraction measures the uncertainty of the initial conditions of the basins and the entropy of the boundary basin ($S_{bb}$). According to [42, 43], the algorithm to determine Sb consists of dividing the basin of attraction into a grid of boxes of linear size $\varepsilon_l$. Each box contains, in principle, infinite initial conditions that generate infinite trajectories, each leading to a labeled color from 1 to $N_A$.

Although $\varepsilon_l$ is our limiting resolution, the information provided by trajectories inside a box can be used to make assumptions about the uncertainty associated with the box. We consider the colors in the box randomly distributed according to some proportions. We can associate a probability to each color j inside a box i as $p_{ij}$ which will be evaluated through computational statistics on the trajectories inside the box. Considering that the trajectories inside a box are statistically independent, the entropy of each box *i* is given by:

$$S_i = \sum_{j=1}^{m_i} p_{i,j} \log\left(\frac{1}{p_{i,j}}\right) \quad (8)$$

Where $m_i \in [1, N_A]$ is the number of colors inside the box *i* and the probability $p_{ij}$ of each color j is determined by the total number of trajectories leading to that color divided by the total number of trajectories in the box. In this way, the total entropy of the entire grid and associated with the N boxes is described by:

$$S = \sum_{i=1}^{N}\sum_{j=1}^{m_i} p_{i,j}\log\left(\frac{1}{p_{i,j}}\right) \qquad (9)$$

We consider the entropy S relative to the total number of boxes N and define the entropy of basins:

$$S_b = \frac{S}{N} \qquad (10)$$

In Figs. (5)(b) and (c) show the behavior of $S_b$ and $S_{bb}$, respectively. It can be observed that the maximum value of $S_b$ and $S_{bb}$ corresponds to $x_3^0 = 0.0$ which shows a greater complexity between the basins of attraction shown in Figs. (4). According to [42] for $S_{bb}$ values above log (2) have a fractal limit, which indicates that it is observed by the uncertainty exponent of the basins of attraction.

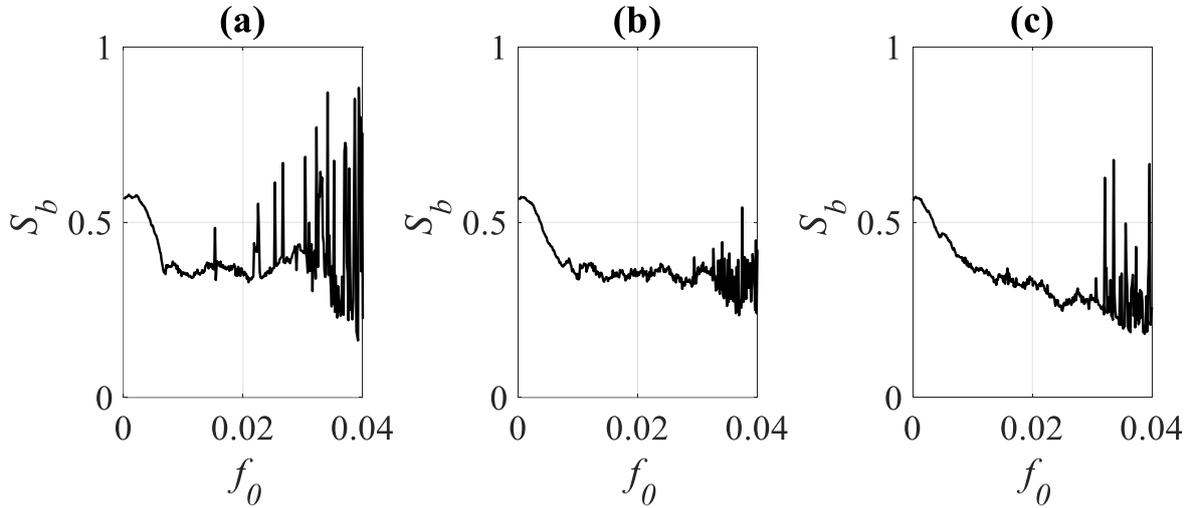

**Fig. (4): Entropy Basins ($S_b$) for (a) $a_0$ = 0.0 and $f_0$ in [0.001, 0.04], (b) $a_0$ = 0.5 and $f_0$ in [0.001, 0.04] and (c) $a_0$ = 1.0 and $f_0$ in [0.001, 0.04].**

In Figures (5) (a)-(c) show the basins of attraction for the minimum values of ($\alpha$) and with their respective values of Sb as shown in Tab. (3).

| $a_0$ | $f_0$ | # attractors | $\alpha$ | $S_b$ | Fig. |
|---|---|---|---|---|---|
| 0.0 | 0.0396 | 5 | 0.0521 | 0.8807 | Fig. (5) (a) |
| 0.5 | 0.0375 | 4 | 0.1928 | 0.5293 | Fig. (5) (b) |
| 1.0 | 0.0336 | 4 | 0.0713 | 0.6642 | Fig. (5) (c) |

**Tab (3): Values of α, $f_0$ and $S_b$ that generate ridding basins and the number of attractors.**

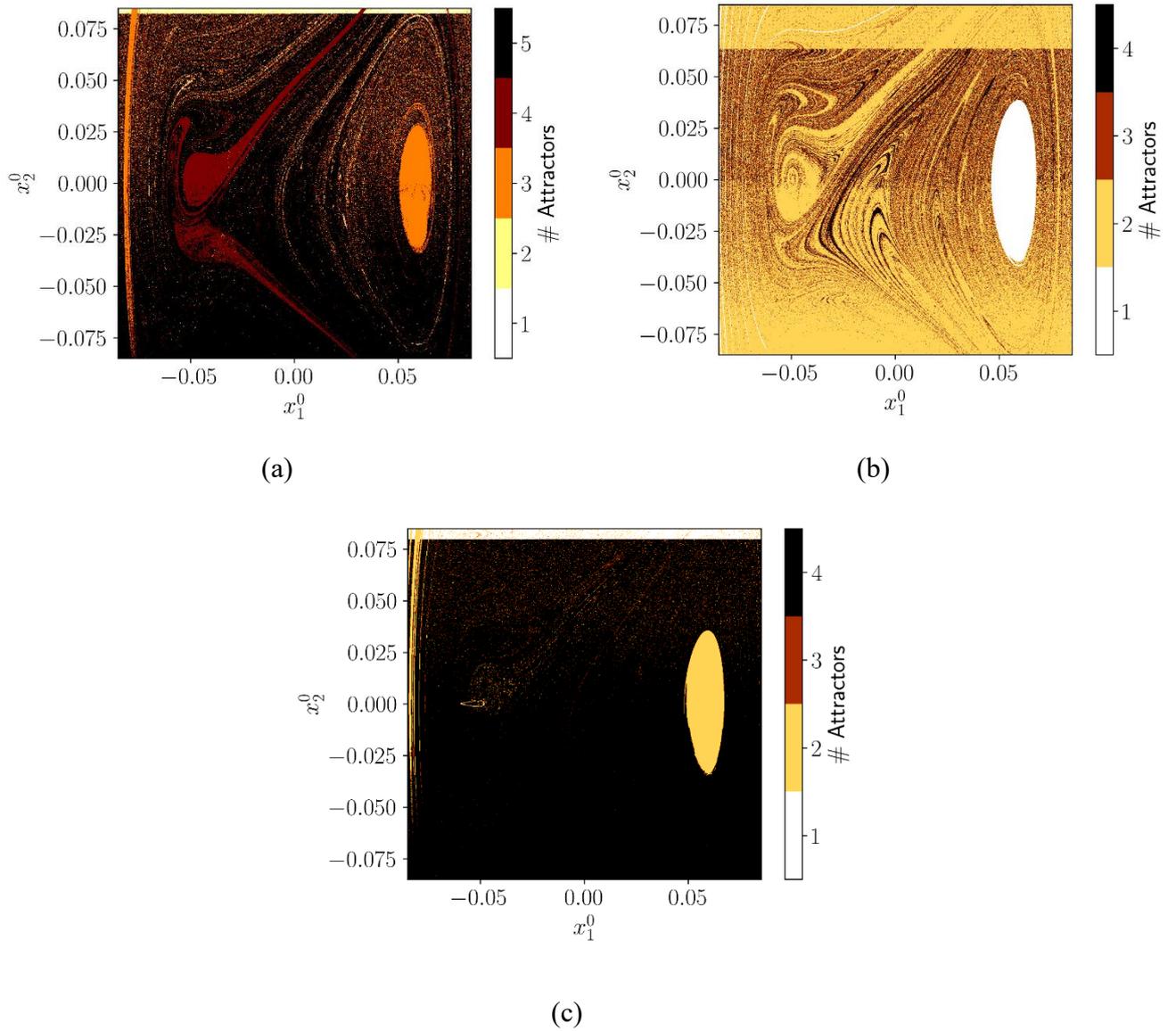

(a)                                                              (b)

(c)

**Fig. (5): Basins of Attraction. (a)** *$a_0 = 0.0$ and $f_0 = 0.0396$*, **(b)** *$a_0 = 0.5$ and $f_0 = 0.0375$* **and (c)** *$a_0 = 1.0$ and $f_0 = 0.0336$*

## 3.2 SPACE OF PARAMETERS ANALYSIS -APPEARANCE OF SHRIMPS.

Therefore, the influence of dimensionless parameters $a_0$ and $f_0$ on the Lyapunov exponent, which are parameters that intervene in the nonlinear dynamics of the system with a non-ideal motor applied to the system. In Figs. (6) show the behavior of the maximum Lyapunov exponent ($\lambda_{max}$), the regions between gray and black show the periodic behavior of the structure ($\lambda_{max} < 0$), however, for colors between1 purple and yellow it shows the chaotic behavior of the structure ($\lambda_{max} > 0$). The Fig (6)(a) is $\lambda_{max}$ for $a_0$ in [0,1] and $f_0$ in [0.001, 0.2] and Fig (6)(b) is the zoom in $a_0$ in [0.71, 0.79] and $f_0$ in [0.1870, 0.2]. The structures formed in the $\lambda_{max}$ parameter space is called shrimps. Structures formed by a regular set of adjacent windows centered around the main pair of super-stable parabolic arches are called prawns. A shrimp is an infinite mosaic of stability domains doubly composed of a main innermost domain and all adjacent stability domains that arise from two period-folding cascades, together with their corresponding chaotic domains.

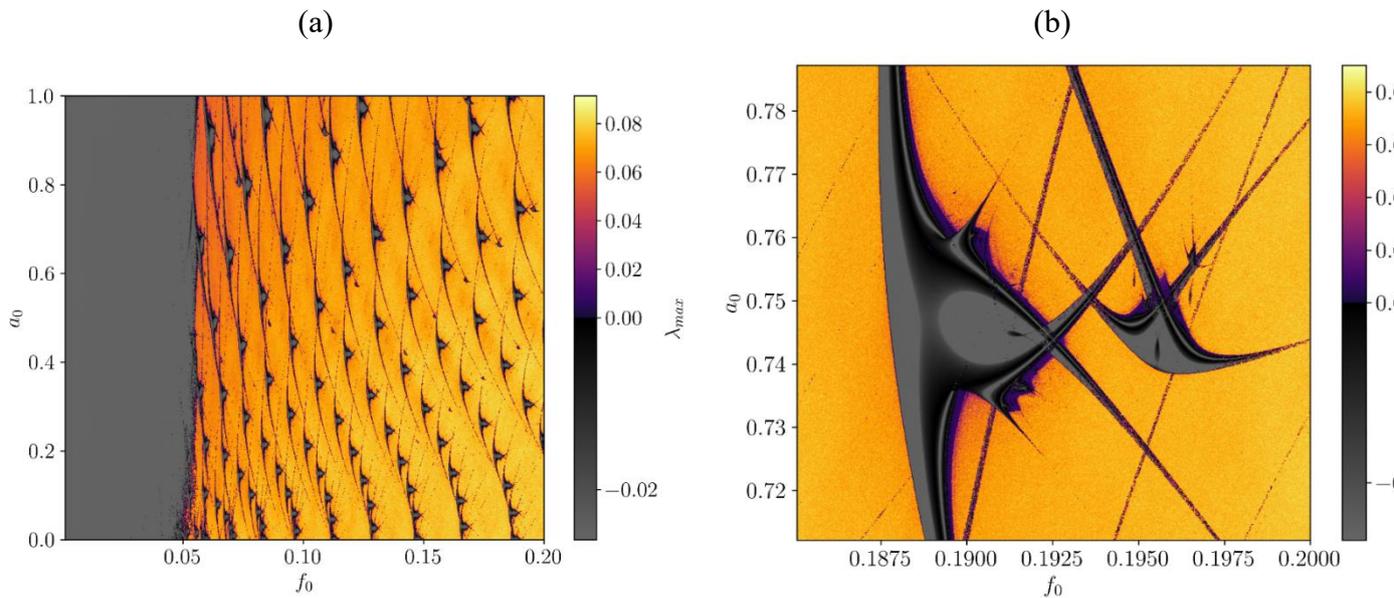

**Fig. (6): $\lambda_{max}$ with $\theta = 0.5$. (a) $a_0$ in [0,1] and $f_0$ in [0.001, 0.2] and (b) The black color is periodic behavior and yellow to red is chaotic behavior.**

With this, we determine the bifurcation diagram for a0=0.75 and $f_0$ in [0, 0.2] in which the periodic windows with their respective periods appear. In Fig(7) shows the behavior of the bifurcation diagram with its corresponding Lyapunov spectrum.

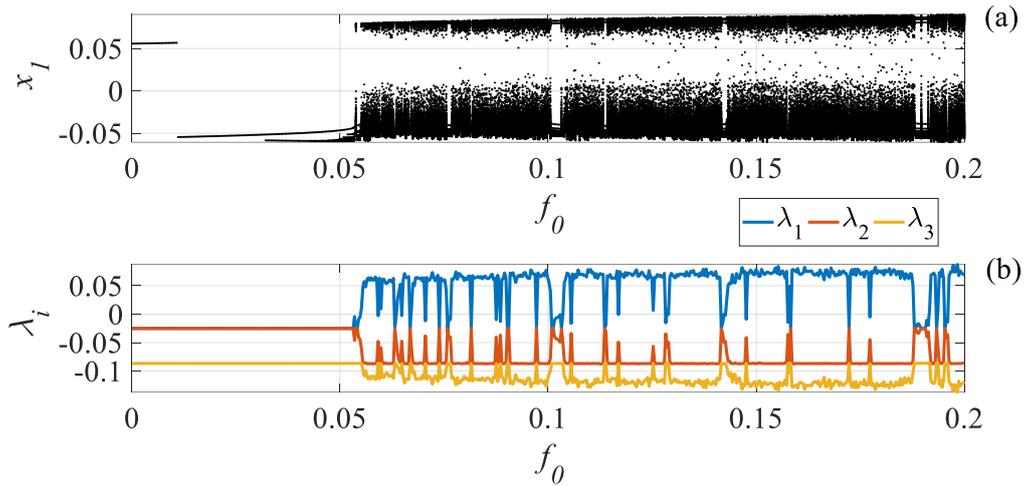

**Figs. (7): (a)** Bifurcation diagram of $x_1$. **(b)** Lyapunov Exponent ($\lambda_i$).

The fig. (8) shows the behavior of the bifurcation diagram with an approximation in $f_0$ in [0.185, 0.2] and $a_0 = 0.75$ that the region is contained in the shrimp structure identified in the Lyapunov exponent in Fig. (8) (b).

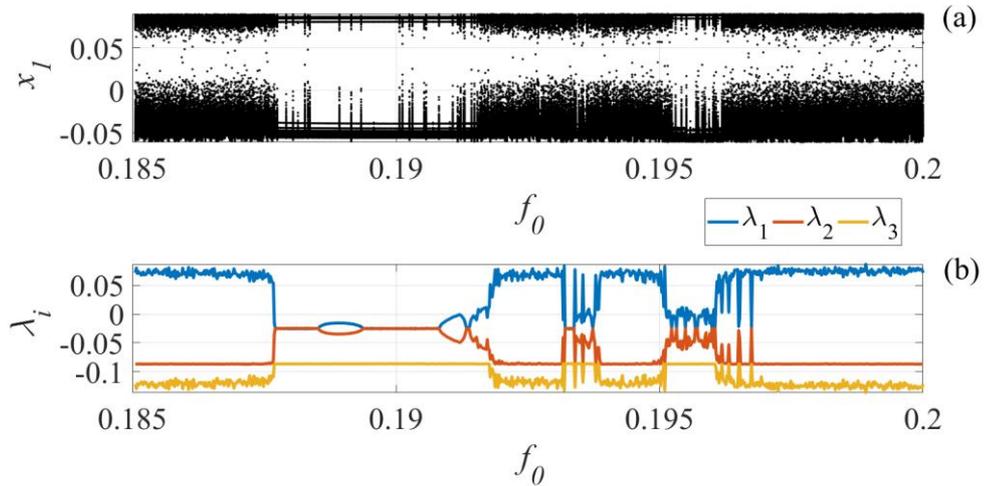

**Figs. (8): (a)** Bifurcation diagram of $x_1$. **(b)** Lyapunov Exponent ($\lambda_i$).

In Tab. (4) shows the intervals of $f_0$ that represent the periodic windows within the interval observed in the bifurcation diagram in fig. (7).

| $a_0 =0.75$ and $f_0$ in [0.0001, 0.2] | | $a_0 =0.75$ and $f_0$ in [0.185, 0.2] | |
|---|---|---|---|
| [0.0001, 0.05483] | See Fig. (7) | [0.1877, 0.1916] | See Fig. (8) |
| [0.07552, 0.07652] | | [0.1932, 0.1934] | |
| [0.09021, 0.09054] | | [0.1934, 0.1938] | |
| [0.1009, 0.1036] | | [0.1951, 0.196] | |
| [0.1136, 0.1142] | | | |
| [0.1282, 0.1289] | | | |
| [0.1413, 0.1429] | | | |
| [0.188, 0.1913] | | | |
| [0.1953, 0.196] | | | |

**Tab. (4): Intervals of the periodic windows of the parameter $f_0$ in [0.0001, 0.2] with $a_0 =0.75$.**

Next, we will consider the phase portraits, for $a_0 = 0.75$ and $f_0 = 0.1212$ chaotic behavior (Fig. (9) (a)) and $f_0 = 0.03214$ is periodic behavior (Fig. (9) (a)), as we may see in the bifurcation diagram Fig (7).

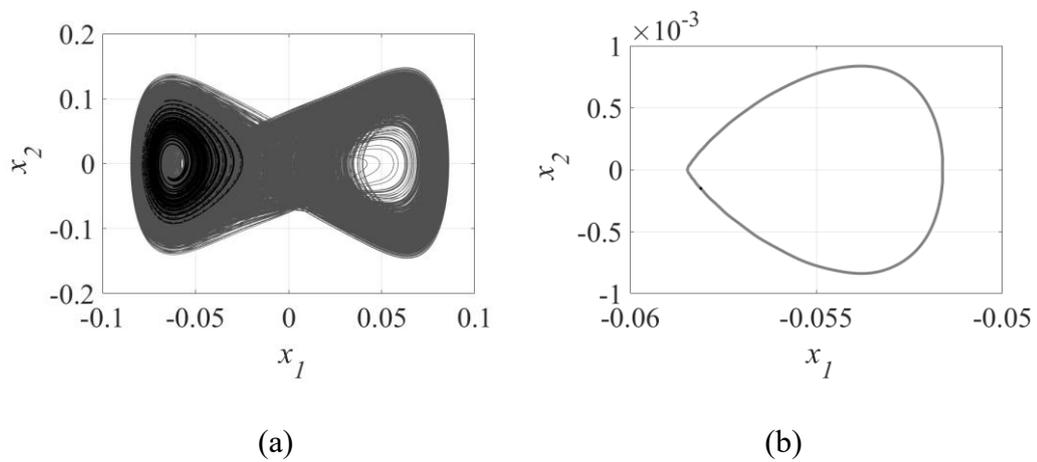

(a)      (b)

**Fig. (9): Portrait phase (gray lines) and Poincare Map (black dots) for (a) $a_0 = 0.75, f_0 = 0.1212$ chaotic behavior and (b) $a_0 = 0.75, f_0 = 0.03214$ periodic.**

## 4. AVERAGE POWER

The average power output was generated by the system of the governing Eqs. (5). According to [10, 44-46], the average power generated can be calculated for:

$$P_{avg} = \frac{1}{T}\int_{t_0}^{T} v^2 d\tau \qquad (8)$$

In Figs. (6) represent the behavior of the average output power calculated by Eqs. (6), we can observe that the regions with the highest average power; however, for the regions in blue, they show the behavior of the minimum average power produced by the device. Determining these regions shows the influence of external force parameters for application in the device in Fig. (1) for energy harvesting. Figs. (6) (a)-(b) show $P_{avg}$ behavior with parametric sweep of parameters $a_0 \times f_0$ respectively.

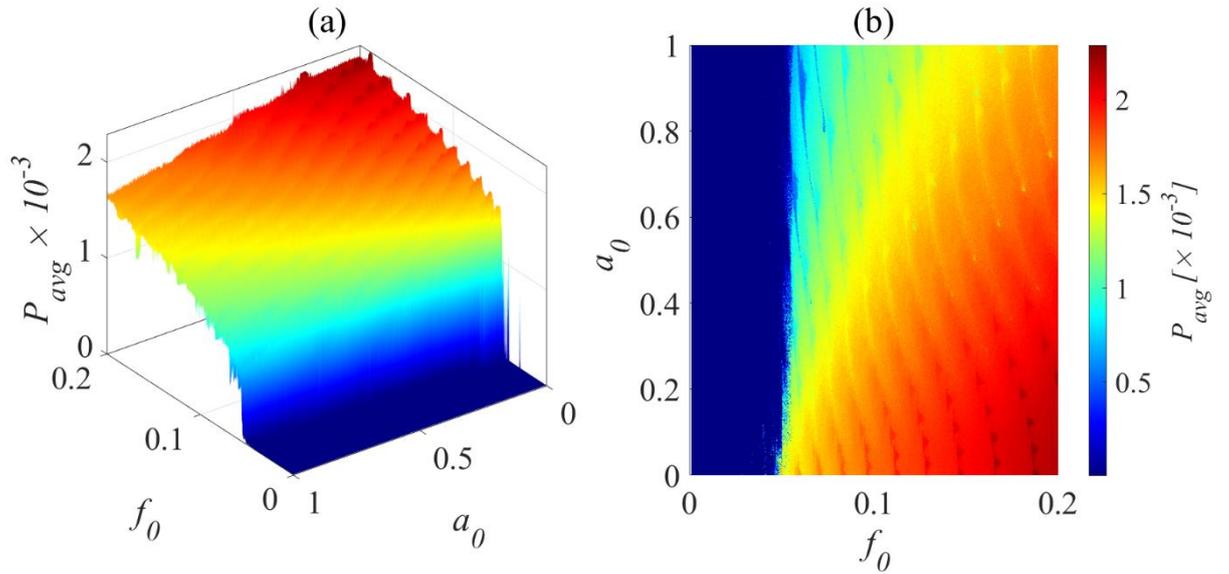

**Fig. (10): Power average output $\theta = 0.5$. (a) $f_0 \in [0, 0.2] \times a_0 \in [0, 1]$ and (b) is the projection to Power average. The blue color is the minimum average power output, and the red is the maximum average power output.**

## CONCLUSIONS

In this work, the nonlinear dynamics of a magnetic system with Shape Memory Alloy (SMA) with polynomial model that receives a non-ideal motor, as an external force, was numerically investigated. The non-ideal motor equation uses its rotation and can be described by two parameters $a_0$, both described in the range of [0,1] and $f_0$ in [0.001, 0.2]. We analyzed the behavior of the initial conditions and observed a degree of fractality. This fractality was observed with the calculation of the uncertainty exponent and $S_b$, as observed in Fig. (3) and Fig (4).

We observed its influence on the nonlinear dynamics of the system, as we can see in Figs. (6) that describe the $\lambda_{max}$. that we obtained in the regions in which the system has chaotic behavior. The changes observed in the scan of the analyzed parameters showed an influence on the average output power because of its coupling to the magnetic mass displacement. This dynamic analysis supports future control designs that suppress chaotic behavior, and thus determine a constant energy production process. Determining the non-linear dynamic behavior of the system for energy harvesting makes it possible to establish parameters that allow the maximum collection of average power and to establish analyzes for future works such as control design for suppression of chaotic motion and with fractional damping with excitation by a non-linear Nonideal motor inspired by [26, 36].

**Funding:**

The authors acknowledge the financial support by the Brazilian Council for Scientific and Technological Development, CNPq,

**Conflicts of Interest:** The authors declare no conflict of interest.